\documentclass[prd,amsfonts,aps,nofootinbib,notitlepage,10pt,superscriptaddress]{revtex4-1}
\pdfoutput=1
\usepackage{amsmath,amssymb}
\usepackage{graphicx}
\usepackage[utf8]{inputenc}
\usepackage{cancel}
\usepackage{color}

\newcommand{\EIAaddress}{\affiliation{Grupo F\'isica Te\'orica y Aplicada, Universidad EIA, A.A. 7516, Medell\'in, Colombia}}
\newcommand{\UdeAaddress}{\affiliation{Instituto de F\'isica, Universidad de Antioquia, Calle 70 No. 52-21, Medell\'in, Colombia}}
\begin{document}
\title{Doublet-triplet dark matter with neutrino masses}
\author{Amalia Betancur}%
\email{amalia@udea.edu.co}
\EIAaddress
\UdeAaddress
\author{Robinson Longas}
\email{robinson.longas@udea.edu.co}
\UdeAaddress
\author{Oscar Zapata}%
\email{oalberto.zapata@udea.edu.co}
\UdeAaddress
\date{\today}

\begin{abstract}
We consider a dark matter (DM) model that arises from the interplay of two renormalizable dark matter models, namely the doublet-triplet fermion model and the doublet-triplet scalar model. 
Despite being excellent exponents of the WIMP paradigm, the physics related to DM in each of these models fails at the same time to account for neutrino masses. It turns out that from the combination of these two models it is possible to generate neutrino masses at one-loop level in the four topologies that are realizations of the Weinberg operator for neutrino masses at one loop. In this work, we combine both models focusing mostly on fermionic dark matter lying at the electroweak scale.  We analyze the impact of the extra charged fields on the Higgs diphoton decay and find that, thanks to the presence of the charged scalars, it is possible to have a viable DM region at the electroweak scale. 
\end{abstract}

\maketitle
\section{Introduction}
\label{sec:introduction}
The astonishing discovery of the Higgs boson at the LHC has closed a stage of the Standard Model (SM) and at the same time has reinforced the proposal of explaining the dark matter (DM) of the Universe by using the Higgs portal ($H^\dagger H$)~\cite{Patt:2006fw}, one of the two dimension-2 terms that are gauge and Lorentz invariant  within the SM. Scalar and vector DM models naturally make use of the Higgs portal, whereas fermion DM models require an ultraviolet realization \cite{Kanemura:2010sh,Lebedev:2011iq,Djouadi:2011aa,Djouadi:2012zc}. 
In particular, we have the very well-known singlet scalar DM model~\cite{Silveira:1985rk,McDonald:1993ex,Burgess:2000yq} as the minimal realization of the Higgs-portal scalar DM, and the singlet fermion~\cite{Kim:2006af,Kim:2008pp,Baek:2011aa}, the singlet-doublet fermion~\cite{ArkaniHamed:2005yv,Mahbubani:2005pt,D'Eramo:2007ga,Enberg:2007rp} and doublet-triplet fermion \cite{Dedes:2014hga} DM models as the ultraviolet realizations of the fermionic Higgs portal.  
Other renormalizable scalar DM models, such as the inert doublet model (IDM)~\cite{Deshpande:1977rw,Barbieri:2006dq} and inert triplet model (ITM)~\cite{Cirelli:2005uq,FileviezPerez:2008bj,Hambye:2009pw},  can also be considered as realizations of the Higgs-portal scalar DM, since the Higgs-mediated scalar interactions also help  the SM yo communicate with the DM particle, leading to processes related to DM annihilation and detection.

In the doublet-triplet fermion dark matter (DTFDM) model \cite{Dedes:2014hga} a vectorlike doublet with $Y=-1$ and a Majorana triplet are added to the Standard Model, both odd under a $Z_2$ symmetry. After the electroweak symmetry breaking, the particle spectrum contains two charged fermions and three Majorana fermions, with the lightest Majorana fermion being the dark matter candidate. The viable dark matter regions are the ones featuring a dark matter mass around the electroweak scale and above 1 TeV \cite{Dedes:2014hga,Abe:2014gua,Freitas:2015hsa}. When the DM particle is mainly doublet (triplet), the correct relic abundance is explained for DM masses around $\sim 1\,(2.8) $ TeV and when the mixing is arbitrary, the abundance can be correctly explained for low masses, $\lesssim 100 $ GeV. 
Regarding the LHC constraints, unlike the high mass region, the low mass region is severely constrained from the Br$(h \rightarrow \gamma  \gamma)$ measurement~\cite{Khachatryan:2016vau}. More specifically, the total $Z_2$-odd fermion contribution to $h \rightarrow \gamma  \gamma$ is always opposite in sign to the SM contribution and sizable due to the large values of the involved Yukawa couplings, which in turn generates a considerable suppression on the Higgs diphoton decay. Since the corresponding decay rate is close to or below the current limit from the ATLAS and CMS data \cite{Khachatryan:2016vau}, we have that the promising low mass region is mostly excluded. 
However, this conclusion can be modified if extra scalar charged particles are added in such a way that the Higgs diphoton decay is altered. 

Despite the fact that the DTFDM model is an excellent exponent of the WIMP paradigm \cite{Jungman:1995df,Bertone:2004pz}\footnote{See Ref. \cite{Arcadi:2017kky} for a recent review.}, the physics associated with the DM sector fails at the same time to account for other evidence of physics beyond the SM, such as neutrino masses \cite{Fukuda:1998mi,Ahmad:2002jz}. Indeed, this fact not only concerns the DTFDM model but all the realizations of the Higgs portal mentioned above, since the total lepton number is always conserved. Nevertheless, if extra scalar degrees of freedom are added to the DTFDM model it is possible to generate neutrino masses through different realizations of the Weinberg operator for neutrino masses at one-loop \cite{Ma:1998dn,Bonnet:2012kz}. For example, if  a $Z_2$-odd doublet scalar with $Y=1$ is added we get the model classified as the T-3-C model with $\alpha=-1$ in Ref. \cite{Restrepo:2013aga}, which allows one-loop neutrino masses through the topology T-3 \cite{Bonnet:2012kz}, as in the radiative type III 
seesaw model \cite{Ma:2008cu,Chao:2012sz,vonderPahlen:2016cbw}.  If instead we add a real triplet scalar, we arrive at a similar conclusion, this time with the model T1-3-H and $\alpha=0$ of Ref. \cite{Restrepo:2013aga} and with the topology T1-III \cite{Bonnet:2012kz}.  
On the other hand, if we add both the doublet and the triplet scalars, the resulting model, which corresponds to the T-1-2-F model with $\alpha=-1$ in Ref. \cite{Restrepo:2013aga}, allows the generation of radiative neutrino masses through four different topologies, namely, T-3, T1-I, T1-II, and T1-III \cite{Bonnet:2012kz}. Consequently this model presents the more complete set of irreducible topologies leading to realizations of the Weinberg operator at one loop \cite{Ma:1998dn,Bonnet:2012kz}, with the interesting feature that all of the $Z_2$-odd fields have an active role in the neutrino mass generation. 

In this paper, we show that such a model, which we denote as the doublet-triplet dark matter (DTDM) model, not only provides a viable fermionic or scalar dark matter candidate, but also constitutes a general framework for radiative neutrino mass generation compatible with the neutrino oscillation data.  Furthermore, when the DM particle is a fermion and lies in the region around the electroweak scale, we show that the model can give rise to the correct Higgs diphoton decay rate~\cite{Khachatryan:2016vau} thanks to the contribution of all the $Z_2$-odd charged particles.

The paper is organized as follows: in Section \ref{sec:mixed} we present the DTDM model and our notation. In Section \ref{sec:DMpheno} we analyze the DM phenomenology of the model, where  we focus mostly on the scenario of fermionic DM around the electroweak scale. Additionally, we  assess the impact of the $Z_2$-odd charged particles on the Higgs diphoton decay in that scenario. 
The one-loop neutrino masses are addressed in Section \ref{sec:neutrinomasses}. Finally, our conclusions are presented in Section \ref{sec:conclusions}.

\section{Doublet-triplet DM model}
\label{sec:mixed}
The model enlarges the fermion sector of the SM by adding an $SU(2)_L$ vectorlike doublet with $Y=-1$ and a Majorana $SU(2)_L$ triplet, both odd under an exact $Z_2$ symmetry, which are expressed as
\begin{align}\label{eq:fermioncontent}
\psi=\left( \begin{array}{ccc}
\psi^0  \\
\psi^- \end{array} \right),\hspace{1cm} 
\Sigma_L=\left( \begin{array}{ccc}
 \Sigma_L^0/\sqrt{2} &  \Sigma_L^+\\
 \Sigma_L^{-} & -\Sigma_L^0/\sqrt{2} \end{array} \right).
\end{align}
On the other hand, the scalar sector is enlarged by two $Z_2$-odd $SU(2)_L$-multiplets, a $Y=1$ doublet and a real triplet: 
\begin{align}\label{eq:scalarcontent}
H_2=\left( \begin{array}{ccc}
H^+  \\
\frac{H^0+i A^0 }{\sqrt{2}} \end{array} \right),\hspace{1cm}
\Delta=\frac{1}{2}\left( \begin{array}{ccc}
 \Delta_0 & \sqrt{2}\ \Delta^+\\
\sqrt{2}\ \Delta^{-} & -\Delta_0 \end{array} \right).
\end{align}
With this additional set of new particles it follows that the most general $Z_2$-invariant Lagrangian of the model can be written as
\begin{align}\label{eq:Ldtdm}
\mathcal{L}&=\mathcal{L}_{\rm{SM}}+\mathcal{L}_{\rm{F}}+\mathcal{L}_{\rm{S}}+\mathcal{L}_{\rm{I}},
\end{align}
where $\mathcal{L}_{\rm{SM}}$ is the SM Lagrangian, which comprises the scalar potential of the Higgs doublet $H_1$, $\mathcal{V}_{\rm{SM}}=-\mu_1^2|H_1|^2+\frac{\lambda_1}{2}|H_1|^4$. Also, $\mathcal{L}_{\rm{F}}$ refers to the kinetic and mass terms of the $Z_2$-odd fermion particles,
\begin{align}
\mathcal{L}_F&=\bar{\psi} i\gamma^\mu D_\mu\psi-M_\psi\bar{\psi}\psi+{\rm Tr}[\bar{\Sigma}_L i\gamma^\mu D_\mu \Sigma_L]-\frac{1}{2}{\rm Tr}(\bar{\Sigma}_L^cM_\Sigma\Sigma_L+\mbox{h.c.}),
\end{align}
whereas $\mathcal{L}_{\rm{S}}$ contains the kinetic, mass, and self-interaction terms of the $Z_2$-odd scalar particles,
\begin{align}
\mathcal{L}_{S}&=|D_\mu H_2|^2-\mu_{2}^2 |H_2|^2 - \frac{\lambda_2}{2} |H_2|^4+{\rm Tr}|D_\mu \Delta|^2 - \mu_{\Delta}^2 {\rm Tr}[\Delta^2] - \frac{\lambda_{\Delta}}{2}{\rm Tr}[\Delta^2]^2.
\end{align}

Lastly, $\mathcal{L}_{\rm{I}}$  contains the different interaction terms between the $Z_2$-odd particles and the SM ones:
\begin{align}\label{eq:LI}
\mathcal{L}_{\rm{I}}= \left[-y_1H_1^\dagger\bar{\Sigma}_L^c\epsilon \psi_R^c   +  y_2 \bar{\psi}_L^c \epsilon \Sigma_L H_1 -\zeta_i\bar{L}_{i}\Sigma_L^c \tilde{H}_2 - \rho_i \bar{\psi}_LH_2 e_{Ri} - f_i \bar{L}_{i}\Delta \psi_R  + {\rm h.c.}\right]-\mathcal{V}_{\rm{I}}.
\end{align}
Here $L_{i}$ and $e_{i}$ represent the SM lepton $SU(2)_L$ doublets and singlets, respectively, and $y_1, y_2, \zeta_i, \rho_i$ and $f_i$ are Yukawa couplings controlling the new interactions ($i=1,2,3$). We assume $M_\psi$ and $y_{1,2}$ to be positive and $M_\Sigma$ to be real \cite{Dedes:2014hga}. The last term in Eq.~(\ref{eq:LI}) accounts for the interaction potential, 
\begin{align}\label{eq:VI}
\mathcal{V}_{\rm{I}}&=\lambda_3|H_1|^2|H_2|^2 + \lambda_4|H_1^{\dagger}H_2|^2  + \frac{\lambda_5}{2} \left[(H_1^{\dagger} H_2)^2 + \rm{h.c.} \right]
 + \lambda'_3|H_1|^2 {\rm Tr} [\Delta^2]+\lambda_{6}|H_2|^2 {\rm Tr}[\Delta^2] + \mu \left[H_1^{\dagger} \Delta H_2 + \rm{h.c.} \right],
\end{align}
where  $\lambda_5$ and $\mu$ have been taken to be real. 
\subsection{Scalar sector}
\label{sec:scalar_sector}
In order to preserve the $Z_2$ symmetry once electroweak symmetry breaking occurs, a zero vacuum expectation value for $H_2$ and $\Delta$ is assumed, along with $\mu_1^2>0$, $\mu_2^2>0$, and $\mu_{\Delta}^2>0$.
This entails that the SM Higgs boson $h$ does not get mixed with the $Z_2$-odd neutral particles and the trilinear $\mu$ term is the unique term responsible for the mixing among the CP-even neutral components as well as the charged components of the doublet and triplet $Z_2$-odd fields. By parametrizing the Higgs doublet as $H_1=(0,\,(h+v)/\sqrt{2})^T$, with $v=246$ GeV, it follows that the CP-even neutral and charged mass matrices in the basis $(H^0, \Delta^0)$ and $(H^\pm, \Delta^\pm)$, respectively, read  
\begin{align}
M_{S^0}=\left( \begin{array}{ccc}
 \mu_2^2+\lambda_Lv^2   & \frac{1}{2}\mu v  \\
 \frac{1}{2}\mu v& \mu_{\Delta}^2 + \frac{1}{2}\lambda'_{3}v^2 \end{array} \right),
\hspace{1cm}
M_{S^\pm}=\left( \begin{array}{ccc}
 \mu_2^2+\frac{1}{2}\lambda_3v^2  & -\frac{1}{2}\mu v \\
-\frac{1}{2}\mu v & \mu_{\Delta}^2 + \frac{1}{2}\lambda'_{3}v^2 \end{array} \right),
\end{align}
where $\lambda_L=(\lambda_3 + \lambda_4 + \lambda_5)/2$ controls the trilinear interaction between the SM Higgs and $H^0$. The CP-even neutral physical states $\eta_{1,2}$ are defined through  
\begin{align}
\left( \begin{array}{cc}
H^0 \\
\Delta^0
\end{array} \right)
= %
\left( \begin{array}{cc}
\cos\alpha & -\sin\alpha \\
\sin\alpha & \cos\alpha
\end{array} \right)
\left( \begin{array}{cc}
\eta_1\\
\eta_2
\end{array} \right), \hspace{2cm}
\sin(2 \alpha)= \frac{ \mu v}{m_{\eta_2}^2-m_{\eta_1}^2}, 
\end{align}
whereas the charged ones $\kappa_{1,2}$ are given by 
\begin{align}
\left( \begin{array}{cc}
H^+ \\
\Delta^+
\end{array} \right)
= %
\left( \begin{array}{cc}
\cos\theta & -\sin\theta \\
\sin\theta & \cos\theta
\end{array} \right)
\left( \begin{array}{cc}
\kappa_1\\
\kappa_2
\end{array} \right),\hspace{2cm}\sin(2 \theta)= \frac{ -\mu v}{m_{\kappa_2}^2-m_{\kappa_1}^2}. 
\end{align}

Finally, $A^0$ remains as the only CP-odd state in the spectrum with $m^2_{A^0}=\mu_2^2+(\lambda_3+\lambda_4-\lambda_5)v^2/2$. 
It follows that the set of free scalar parameters of this model is chosen to be $ m_{A^0}, m_{\kappa_1}, m_{\eta_1},m_{\eta_2},\lambda_2, \lambda_\Delta, \lambda_3, \lambda'_3, \lambda_{6}$, and $\mu$, with the quartic couplings subject to the following vacuum stability and perturbativity conditions \cite{Barbieri:2006dq,Hambye:2009pw,Ilnicka:2015jba}:

\begin{align}
 \label{eq:conditionVpositive3}
 & \lambda_3 + \sqrt{\lambda_1\lambda_2}>0;\; \lambda_3 + \lambda_4 - |\lambda_5| + \sqrt{\lambda_1\lambda_2} >0;\;\lambda_3' + \sqrt{\lambda_1\lambda_\Delta}>0;\;
\nonumber\\ 
 & \lambda_{6} + \sqrt{\lambda_2\lambda_\Delta}>0;\; \lambda_{3},\lambda'_3, \lambda_{6}<4\pi;\; \lambda_{2}, \lambda_{\Delta}<\frac{4\pi}{3}.
\end{align}
Thus we expect either $\eta_1$, $\eta_2$, or $A^0$ to be the lightest particle in the $Z_2$-odd scalar spectrum for an appropriate choice of the scalar couplings.  

\subsection{Fermion sector}
\label{sec:fermionsector}
Since $H_1$ is the only scalar having a nonzero vacuum expectation value (VEV), the $y_{1,2}$ terms in Eq.~(\ref{eq:LI}) are the only ones that generate a mixing between $\psi$ and $\Sigma_L$. The $\zeta_i, \rho_i$, and $f_i$ terms  represent pure interaction terms that may induce both coannihilation and lepton flavor violation (LFV) processes, and two of them ($\zeta_i$ and $f_i$) enter in the neutrino mass generation. 
Consequently, the $Z_2$-odd fermion spectrum of the DTDM model is the one of the DTFDM model \cite{Dedes:2014hga}.    
Thus, the fermion mass matrices  for the neutral [in the basis $\Xi^0=(\Sigma_L^0, \psi^0_L, \psi^{0c}_R)^T$] and charged [in the basis $\Xi^-_R=(\Sigma^{+c}_L, \psi_R^{-})^T$ and $\Xi^-_L=(\Sigma^{-}_L,\psi^-_L)^T$] sectors are given by

\begin{align}
\label{eq:Mchi}
  \mathbf{M}_{\Xi^0}=\begin{pmatrix}
 M_\Sigma                 &\frac{1}{\sqrt{2}}yv\cos\beta& \frac{1}{\sqrt{2}}yv\sin\beta\\
\frac{1}{\sqrt{2}}yv\cos\beta &  0                  & M_\psi\\
\frac{1}{\sqrt{2}}yv\sin\beta&  M_\psi                &  0  \\
\end{pmatrix},\hspace{1cm}
  \mathbf{M}_{\Xi^\pm}=\begin{pmatrix}
 M_\Sigma         &   yv\cos\beta \\
 yv\sin\beta & M_\psi \\
\end{pmatrix},
\end{align}
with $y=\sqrt{(y_1^2 + y_2^2)/2}$ and $\tan\beta=y_2/y_1$\footnote{Note that these mass matrices are reminiscent of the very well-known neutralino and chargino mass matrices (in the decoupled bino limit) in the minimal supersymmetric standard model \cite{Martin:1997ns}.}. 

It follows that the $Z_2$-odd particle spectrum of this model includes two charged fermion particles $\chi_{1,2}^\pm$ with masses $m_{\chi_{1,2}^\pm}= \frac{1}{2}\left[M_{\psi} + M_{\Sigma} \mp \sqrt{(M_{\psi}-M_{\Sigma})^2 + 2y^2v^2}\right]$ implying that $m_{\chi_2^\pm}>m_{\chi_1^\pm}$, and three neutral Majorana states, $\chi_1^0$, $\chi_2^0$ and $\chi_3^0$ (not mass ordering is implied), 
with masses ruled by the characteristic equation
\begin{align}\label{eq:eigvals}
&(M_\Sigma-m_{\chi_i^0})(m_{\chi_i^0}^2-M_\psi^2)+\frac{1}{2}y^2v^2(M_\psi\sin2\beta+m_{\chi_i^0})=0. 
\end{align} 
Clearly the $Z_2$-odd physical states are an admixture of the triplet and two doublets, with nonzero couplings to the $Z$ and Higgs bosons. On the contrary, in the symmetric case $y_1=y_2$ ($\tan\beta=1$) one of the neutral states is an equal admixture of the doublet fermions without a triplet component and does not get a mass from the electroweak symmetry breaking. This means that the neutral spectrum has one pure doublet state with a mass given by the vectorlike mass. 
This can be easily understood after considering the similarity transformation  $\mathbf{M}'_{\Xi^0}=O^\dagger\mathbf{M}_{\Xi^0}O$, with
\begin{align}\label{eq:Mchi2}
\mathbf{M}'_{\Xi^0}=\begin{pmatrix}
 M_\Sigma                 &yv& 0\\
yv &  M_\psi                  & 0\\
0 &  0                &  -M_\psi  \\
\end{pmatrix},
\hspace{1cm}{\rm and}\hspace{1cm}
 O=\begin{pmatrix}
 1  & 0 & 0\\
0 & \frac{1}{\sqrt{2}} & -\frac{1}{\sqrt{2}} \\
0 &\frac{1}{\sqrt{2}} & \frac{1}{\sqrt{2}} \\
\end{pmatrix}.
\end{align}
Thus, we have that $m_{\chi_1^0}=M_\psi$ and the charged eigenstates are degenerate with the other two neutral states, i.e. $m_{\chi_2^0}= m_{\chi_1^{\pm}}$ and $m_{\chi_3^0}= m_{\chi_2^{\pm}}$\footnote{Note that the charged states (and therefore $\chi_2^0$ and $\chi_3^0$) are degenerate when  
\begin{align} \label{eq:condition1}
M_\Sigma=-M_\psi.
\end{align}
That is, all the fermion spectrum but $\chi_1^0$ is degenerate with a mass given by $M_\psi\left[(1+y^2v^2)/(2M_\psi)\right]^{1/2}$. 
}. 
In addition to this, the fermion sector also presents other interesting features in the symmetric case~\cite{Dedes:2014hga}. First, there is a global $SU(2)_R$ symmetry that guarantees a null contribution to the electroweak $T$ parameter and all the mass eigenstates have no diagonal tree-level couplings to the $Z$ boson, i.e., $g_{\chi_i\chi_i Z}=0$. 
Second, when the condition $M_\Sigma<(y^2v^2-4M_\psi^2)/(4M_\psi)$ is fulfilled (and all the $Z_2$-odd scalars are heavier than $\chi_1^0$), the resulting DM candidate is pure doublet ($|m_{\chi_1^0}|<|m_{\chi_2^0}|,|m_{\chi_3^0}|$) with a vanishing diagonal coupling to the Higgs boson $g_{\chi_1\chi_1 h}$ at tree level. 
 
These features have a profound impact on DM phenomenology analysis when $\chi_1^0$ is the DM candidate since the direct detection via the exchange of a $Z$-boson or a Higgs boson would be zero at tree level  (see Sec. \ref{sec:DMpheno}).

\section{DM phenomenology}
\label{sec:DMpheno}

If the lightest $Z_2$-odd particle is electrically neutral, either a fermion or a scalar, it will play the role of the DM particle. 
Consequently, two main scenarios emerge according to whether the $\zeta_i, \rho_i, f_i$ interactions affect the DM annihilation or not. 
When these interactions are suppressed, the resulting DM phenomenology will be very much like that of the doublet-triplet fermion (doublet-triplet scalar) DM model if the DM particle is a fermion (scalar), with the $Z_2$-odd scalars (fermions) not playing any role in the DM annihilation and detection processes. 
On the other hand, to be efficient such interactions demand large values ($\gtrsim1$) for the Yukawa couplings and/or that the $Z_2$-odd scalar and fermion sectors be mass degenerate. In what follows we will only consider the scenario where the $\zeta_i, \rho_i, f_i$ mediated processes are not taking part in the DM annihilation, which is in turn favored by the  bounds coming from lepton flavor processes which, in general, favor small Yukawa couplings.  
\subsection{Fermion DM}
\begin{figure}[t]
\begin{center}
\includegraphics[scale=0.5]{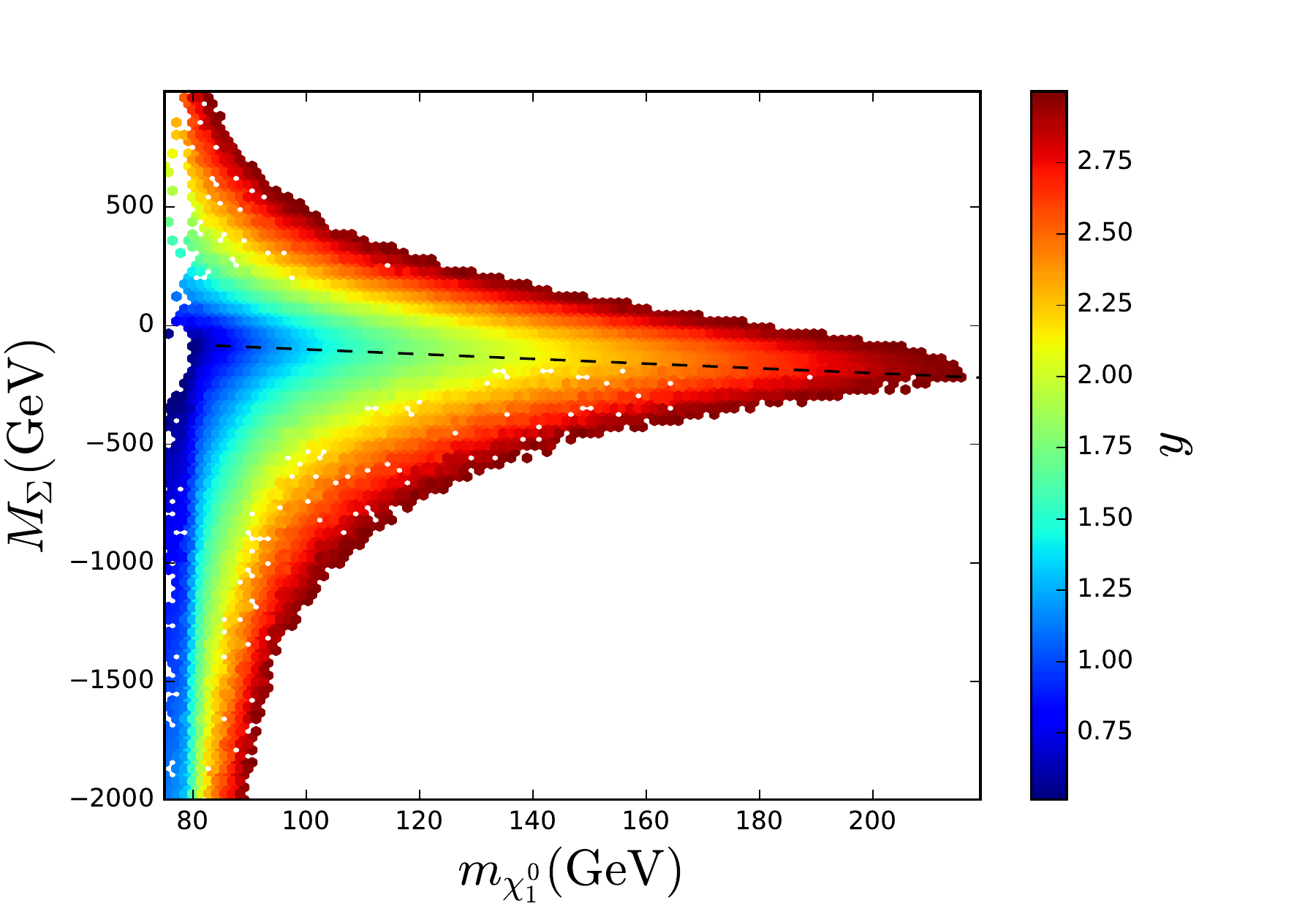}
\caption{The triplet fermion mass as a function of the dark matter mass in the electroweak DM region. The color code denotes the allowed values of $y$ and the dashed line corresponds to the points satisfying $M_\Sigma=-m_{\chi_1^0}$. }
\label{fig:Omega}
\end{center}
\end{figure}
When the DM particle is a fermion, the model can account for the observed DM relic abundance in two distinct regions \cite{Dedes:2014hga,Abe:2014gua,Freitas:2015hsa}: one where the DM candidate lies at the electroweak scale and one around the TeV scale.

The heavy DM scenario arises when the fermion mixing terms are small $y_1,y_2\ll1$ and the DM particle is mainly either doublet or triplet. Since the tree-level mass degeneracy between the DM particle and its charged $SU(2)_L$ partner is lifted radiatively, and is thus small, the coannihilations are very efficient and the observed DM relic abundance is only satisfied for heavy DM around the TeV scale.  If the DM is mostly doublet, the resulting model is similar to the pure doublet fermion DM and the lowest mass to saturate the relic abundance is $\sim 1$ TeV \cite{Chattopadhyay:2005mv,Mahbubani:2005pt,Cheung:2013dua}, whereas, for mostly triplet, the lowest mass is $\sim 2.8$ TeV \cite{Cirelli:2005uq,Chao:2012sz}.

On the other hand, the electroweak DM region results when the DM particle is $\chi_1^0$ in the symmetric case. Since its diagonal couplings to the $Z$ and Higgs bosons vanish at tree level, the DM does not annihilate through the $s$-channel, but only through the $t$- and $u$-channels into gauge bosons $W^+ W^-$ and $Z Z$ via the exchange of one of the $Z_2$-odd heavier fermion eigenstates. The annihilation channels are thus suppressed, and the total relic abundance is achieved for low DM masses, between 80 and 220 GeV, and a large Yukawa coupling, $y \gtrsim 1$. To understand this last requirement, we may take a look at the expression for $m_{\chi_{1,2}^\pm}$, if $y$ is small, one of the heavier masses will be nearly degenerate with the DM, thus making the annihilations very efficient. This possibility may be considered in a way similar to that of the doublet fermion dark matter, which requires masses around the TeV scale \cite{Chattopadhyay:2005mv,Mahbubani:2005pt,Cheung:2013dua}, but this is far from the much more 
appealing electroweak region. On the other hand, if $y \sim 1$, the splitting between the neutral eigenstates is large, annihilation is less efficient, and relic abundance is saturated at masses around 100 GeV. 
In Figure \ref{fig:Omega} we display the resulting parameter space consistent with the DM relic abundance reported by the Planck Collaboration~\cite{Ade:2013lta} after scanning over the free parameters of the model\footnote{The model was implemented in SARAH-4.4.2 \cite{Staub:2013tta} which generates an output to SPheno \cite{Porod:2003um,Porod:2011nf} to obtain the physical spectrum, which was then exported to MicrOMEGAS \cite{Belanger:2013oya} to calculate the relic abundance.}. 
The maximum value for $y$ is set as a perturbativity condition. Below the minimum value for $m_{\chi_1^0}\sim 80$ GeV, DM annihilations are no longer efficient\footnote{To be consistent with collider bounds we have demanded that the lightest charged fermion is heavier than $93$ GeV \cite{Freitas:2015hsa}.} while beyond $\sim210$ GeV they are too efficient. 
Note also that for any value of the pair $y$-$m_{\chi_1^0}$ that saturates relic density, there are two different allowed triplet mass regions: one where $M_{\Sigma}$ is always negative ($M_{\Sigma}<-m_{\chi_1^0}$) and one where it can be either positive or negative but larger than $-m_{\chi_1^0}$. 

In regards to direct detection for this scenario, the dispersion with nuclei is not possible at tree level because spin-independent interactions rely on $Z$ and/or $h$ mediation and neither of them is present at leading order.  However, spin-independent interactions are allowed through loops mediated by the heavier $Z_2$-odd fermions and gauge bosons and through box (and twisted) diagrams involving gauge bosons. 
In principle, the loop suppression could take the cross section out of reach of the sensitivity of current experiments, but due to the large Yukawa couplings required by the relic abundance constraint, the LUX experiment \cite{Akerib:2012ys} does place constraints on a portion of the parameter space, as shown in Ref. \cite{Freitas:2015hsa}.

Since there are two allowed triplet mass regions, direct detection places different constraints on each region. To analyze them, we computed the effective $h-\chi_{1}^0-\chi_{1}^0$ coupling using the expressions given in Ref. \cite{Freitas:2015hsa}; we then calculated the spin-independent cross section, and imposed the most recent bound reported by the LUX experiment \cite{Akerib:2016vxi}. 

\begin{figure}[t!]
\begin{center}
\includegraphics[scale=0.48]{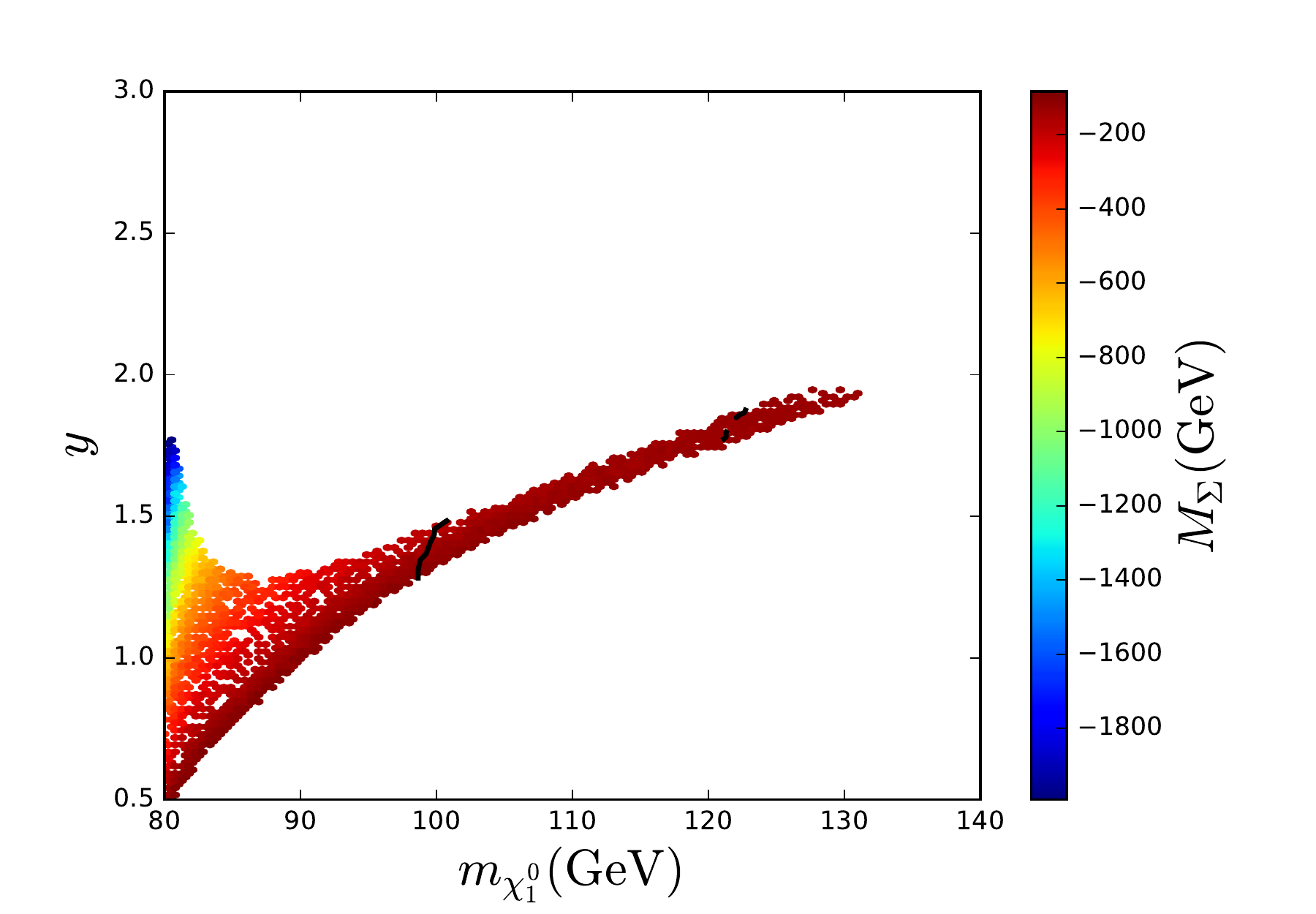}
\includegraphics[scale=0.48]{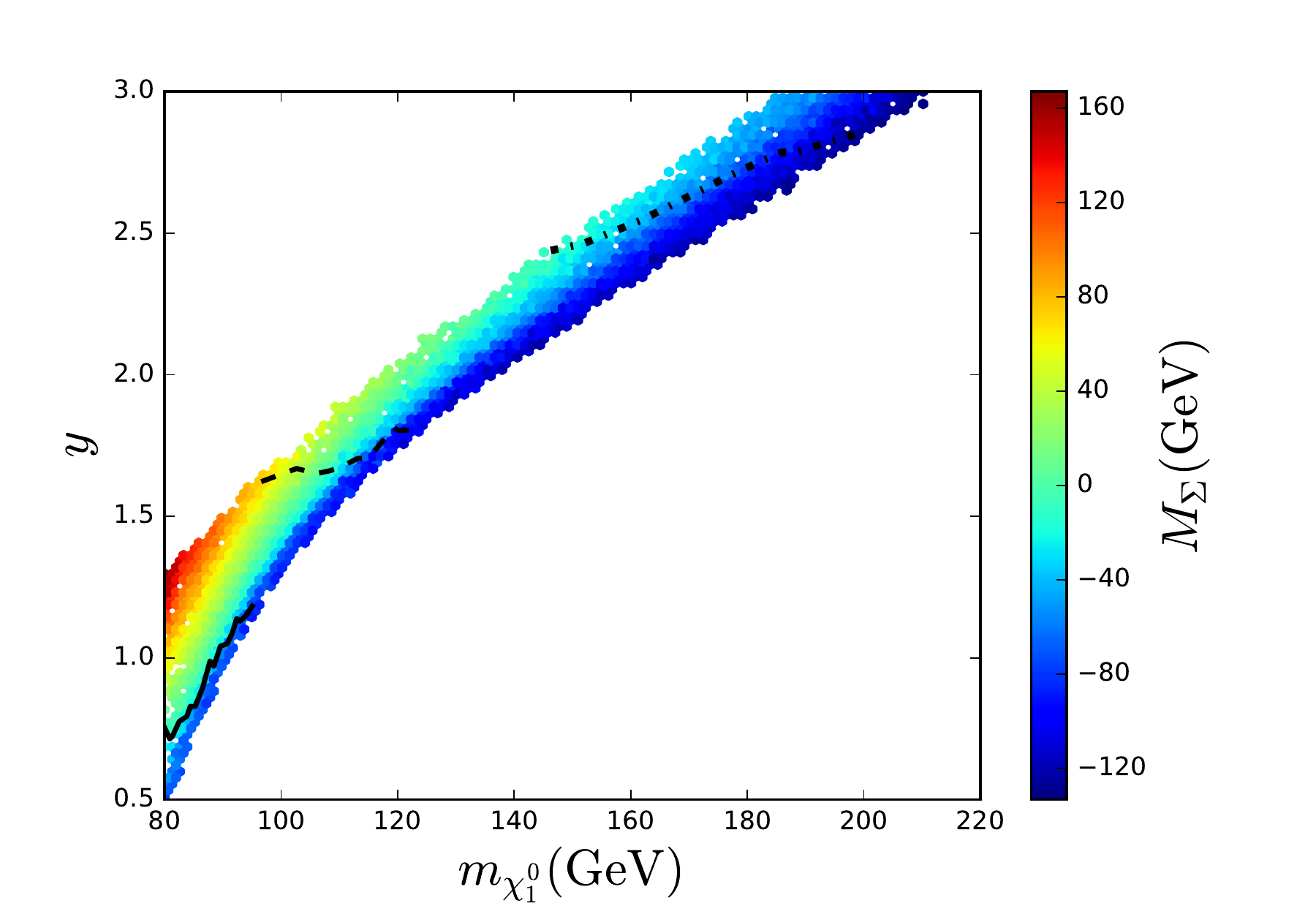}
\qquad
\includegraphics[scale=0.48]{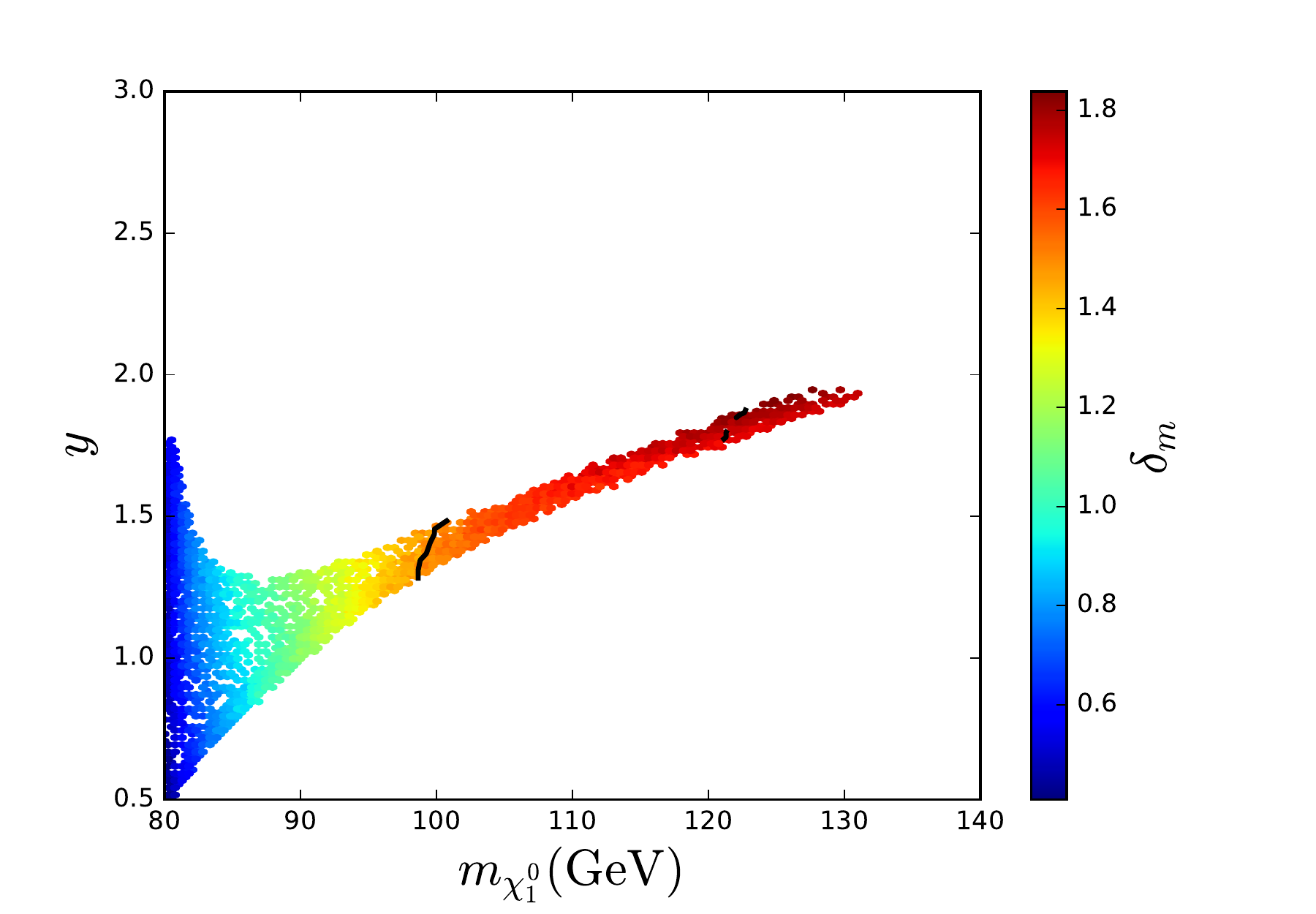}
\includegraphics[scale=0.48]{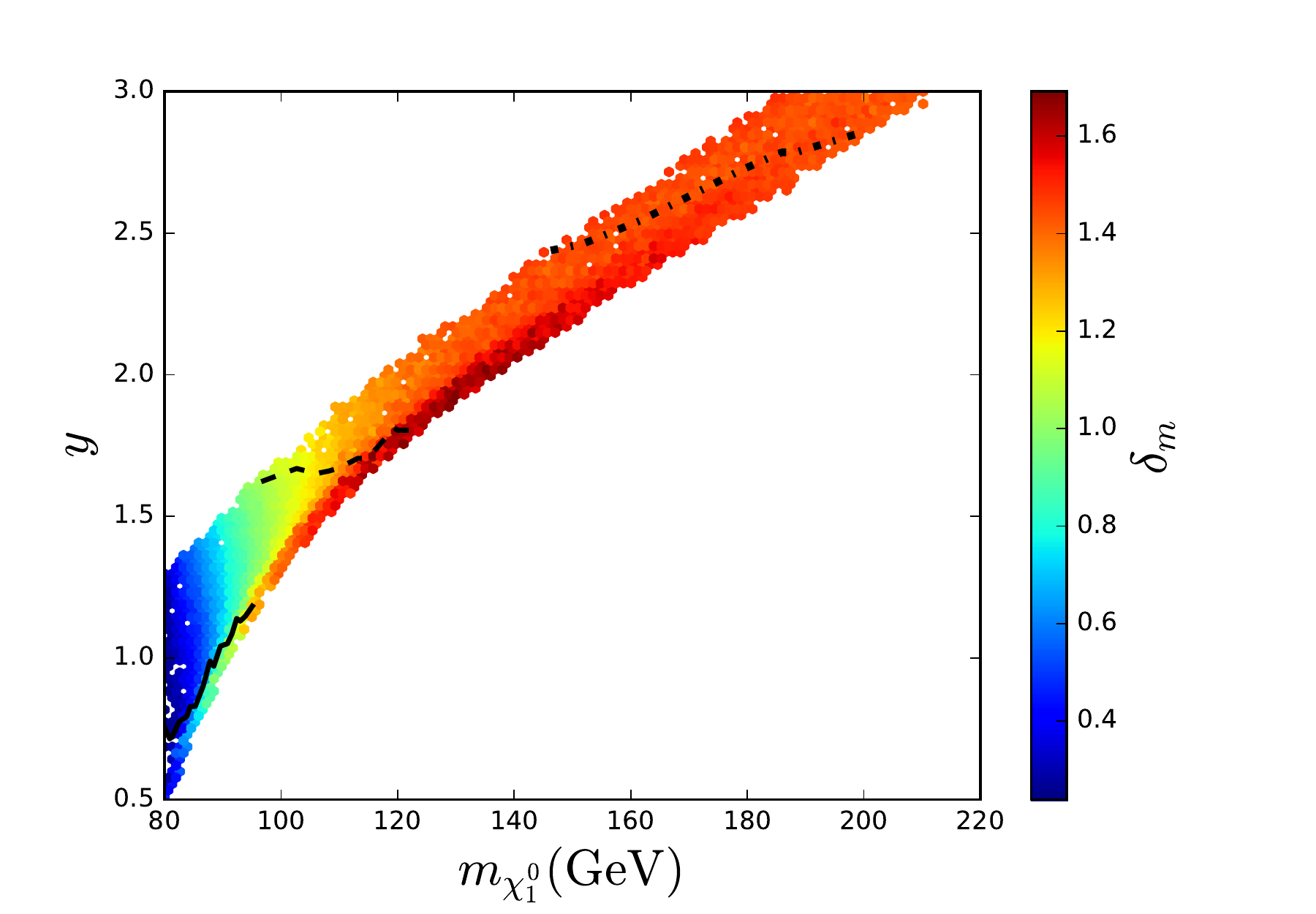}
\qquad
\caption{Parameter space of the electroweak DM region accounting for the observed DM relic abundance and consistent with direct and indirect searches of DM. In the top row the color bar corresponds to the allowed values of $M_{\Sigma}$ while in the bottom row it corresponds to the allowed values of $\delta_m$, as defined in the text. The region in the left (right) panels satisfies $M_{\Sigma} < -m_{\chi_1^0}$ ($M_{\Sigma} > -m_{\chi_1^0}$), which corresponds to the region below (above) the dashed line in Fig.~\ref{fig:Omega}. Solid, dashed and dotted-dashed lines represent the maximum value for $m_{\chi_1^0}$ consistent with the Higgs diphoton decay rate reported in Refs.~\cite{Khachatryan:2016vau}, \cite{CMS:2016ixj} and \cite{ATLAS:2016nke}, respectively (see text for details).} 
\label{fig:lux}
\end{center}
\end{figure}

The top row of figure \ref{fig:lux} shows the allowed parameter space in the $y-m_{\chi_1^0}$ plane with the color bar representing the triplet mass $M_{\Sigma}$. In the region where $M_{\Sigma}\lesssim -m_{\chi_1^0}$ (left panel), for the lowest $m_{\chi_1^0}$, $M_{\Sigma}$ can be as low as $\sim -1900$ GeV, however, as $m_{\chi_1^0}$ increases, $M_{\Sigma}$ becomes heavily restricted very rapidly, to the point that, for $m_{\chi_1^0}$ around $85$ GeV $M_{\Sigma}$  cannot be less than $\sim -500$ GeV. Upon further increasing $m_{\chi_1^0}$, direct detection becomes even more restrictive, leaving only a narrow strip in the $y$ vs $m_{\chi_1^0}$ plane and with the largest $m_{\chi_1^0}$ being less 
than 130 GeV, at that point, $m_{\chi_1^0} \sim -M_{\Sigma}$. On the other hand, in the region where $M_\Sigma\gtrsim -m_{\chi_1^0}$ (right panel), at low values of $m_{\chi_1^0}$, direct detection restricts $M_\Sigma$ to be less than $\sim 160$ GeV. As the dark matter mass increases, negative triplet masses are favored, and for the largest DM mass considered it follows that $M_{\Sigma}\sim-120$ GeV. Notice that in this region all values of the Yukawa coupling considered ($0.5\leq y\leq 3$) are still viable, in contrast to the former region where $y < 1.8$. 
The direct detection bounds also imply that the splitting between the DM mass and the next heavier fermion are also constrained by LUX \cite{Akerib:2016vxi}. Defining $\delta_{m}$ as $(|m_{\chi_2^0}|-|m_{\chi_1^0}|)/|m_{\chi_1^0}|$ we find that, according to the lower right panel of figure \ref{fig:lux}, in the region where $M_{\Sigma}$ is mostly positive, small $\delta_m$ are only allowed for low DM mass and low $y$; {\it e.g.}, for $\delta_{m}<1$ the DM mass must be less than $110$ GeV and $y$ less than $1.5$. 
As the DM mass increases, so must $\delta_{m}$, with the largest allowed value being $\sim 1.6$. A similar situation arises in the region where $M_{\Sigma}$ is always negative. For $\delta_m <1$ the dark matter mass must be less than 90 GeV; a larger DM mass requires greater $\delta_m$, with the largest allowed value being 1.8, as shown in the lower left panel of figure \ref{fig:lux}. It is also worth noting that the color bar for both cases shows that coannihilations between the DM and any of the heavier $Z_2$-odd fermions are not allowed. The lowest allowed $\delta_m$ is $\sim 0.23$ which occurs in the region where  $M_\Sigma>-m_{\chi_1^0}$  (right bottom panel) hence, coannihilations are always Boltzmann-suppressed. This constraint on the splitting comes from relic density, if coannihilations are possible dark matter is more efficiently depleted in the early universe. Such a scenario demands multi-component dark matter. 

Indirect detection can also provide valuable information about the parameter space of the model. In particular, the null observation of the gamma-ray excess in dwarf spheroidal galaxies (dSphs) places constraints \cite{Ackermann:2015zua} on the thermally averaged cross section $\langle \sigma  v \rangle $. In the same vein, the constraints \cite{Ibarra:2013zia,Giesen:2015ufa} from the positron and antiproton measurements \cite{Aguilar:2013qda,Aguilar:2016kjl} are also relevant. 
In the case of the DTDM model, for the region of interest, the production of secondary cosmic rays arises from dark matter annihilation into $W^+ W^-$ and $Z Z$ via the $t$- and $u$-channels. We have compared this to the aforementioned studies and found that since $\langle \sigma  v \rangle $ is at most $\sim 2\times 10^{-26}$ $\rm{cm^3/s}$, it always lies well below the excluded regions. As a result, indirect detection does not place any additional constraint on the parameter space of the model.

Despite being an appealing and promising scenario, the electroweak DM region is severely constrained by the LHC Run-1 measurement of the Higgs diphoton decay rate \cite{Khachatryan:2016vau} since the $Z_2$-odd charged states induce the Higgs decay to photons at one loop.
In fact, in the DTFDM model, almost the whole parameter space of this scenario is excluded\footnote{It is worth mentioning that the preliminary Run-2 limits reported  independently  by CMS~\cite{CMS:2016ixj} and ATLAS~\cite{ATLAS:2016nke} are below the limit from the Run-1 combined analysis~\cite{Khachatryan:2016vau}. In particular, considering only the ATLAS limit~\cite{ATLAS:2016nke} the allowed region slightly increases: $m_{\chi_1^0}\lesssim110$ GeV for $M_\Sigma<-m_{\chi_1^0}$ and $m_{\chi_1^0}\lesssim95$ GeV for $M_\Sigma>-m_{\chi_1^0}$.} due to the large suppression on $R_{\gamma\gamma}$ induced by the two $Z_2$-odd charged fermions (such a suppression arises because the fermion contribution is always positive, that is, opposite in sign to the SM contribution, and sizable due to the large values of $y$). 

Thus, in principle the same should occur in the DTDM model\footnote{The DM region above TeV scale does not suffer from such difficulties since the $Z_2$-odd charged fermions lie above the TeV scale and therefore their effect on the Higgs diphoton decay rate is negligible.}.  However, in the DTDM model there are extra charged scalar fields, $\kappa_{1,2}$, that also mediate the Higgs decay to two photons at one loop and may help to increase $R_{\gamma\gamma}$.
To investigate the impact of the new $Z_2$-odd charged scalar, we consider the limit where $\mu\ll v$, which is favored by electroweak precision observables. The corresponding decay ratio reads \cite{Dedes:2014hga,Freitas:2015hsa,Abe:2014gua}
\begin{align}
\label{eq:higgsgammagamma}
&R_{\gamma\gamma}= \left|1 + \frac{1}{A_{SM}}\left[\frac{\lambda_3\, A_S(\tau_{\kappa_1})}{4 m^2_{\kappa_1}} +  \frac{\lambda_3'\, A_S(\tau_{\kappa_2})}{4 m^2_{\kappa_2}} + \frac{y^2v^2}{m_{\chi_2^\pm}-m_{\chi_1^\pm}}\left(\frac{A_F(\tau_{\chi_2^\pm})}{m_{\chi_{2}^{\pm}}}-\frac{A_F(\tau_{\chi_1^\pm})}{m_{\chi_{1}^{\pm}}}\right)\right]  \right|^2,
\end{align}
where $A_{{\rm SM}}=-6.5$ is the SM contribution from charged fermions and gauge bosons, $A_F(\tau)= 2 \tau ^{-2}[\tau + (\tau-1)\arcsin^2{\sqrt{\tau}}]$ and  $A_S(\tau)=-\tau^{-2}(\tau-f(\tau))$ for $\tau\leq1$,  and $\tau_X=m^2_{h}/(4 m^2_{X})$. It turns out that there are two possibilities in which the scalar contribution may modify the ratio to yield a result that is in agreement with experimental measurements:
$i)$ a sufficiently negative $\lambda_3$,$\lambda_3'$ that counteract the positive contribution from the $Z_2$-odd fermions, and $ii)$ a sufficiently positive $\lambda_3$,$\lambda_3'$ that generate a negative but large contribution (twice that of the SM contribution). 
Both possibilities are subject to the requirement of vacuum stability and perturbativity Eq.~(\ref{eq:conditionVpositive3}) which demand $\lambda_3$, $\lambda_3' \gtrsim -1$ and $\lambda_3$, $\lambda_3' \lesssim 9$, thus leaving the former possibility as the viable one.

\begin{figure}[t!]
\begin{center}
\includegraphics[scale=0.48]{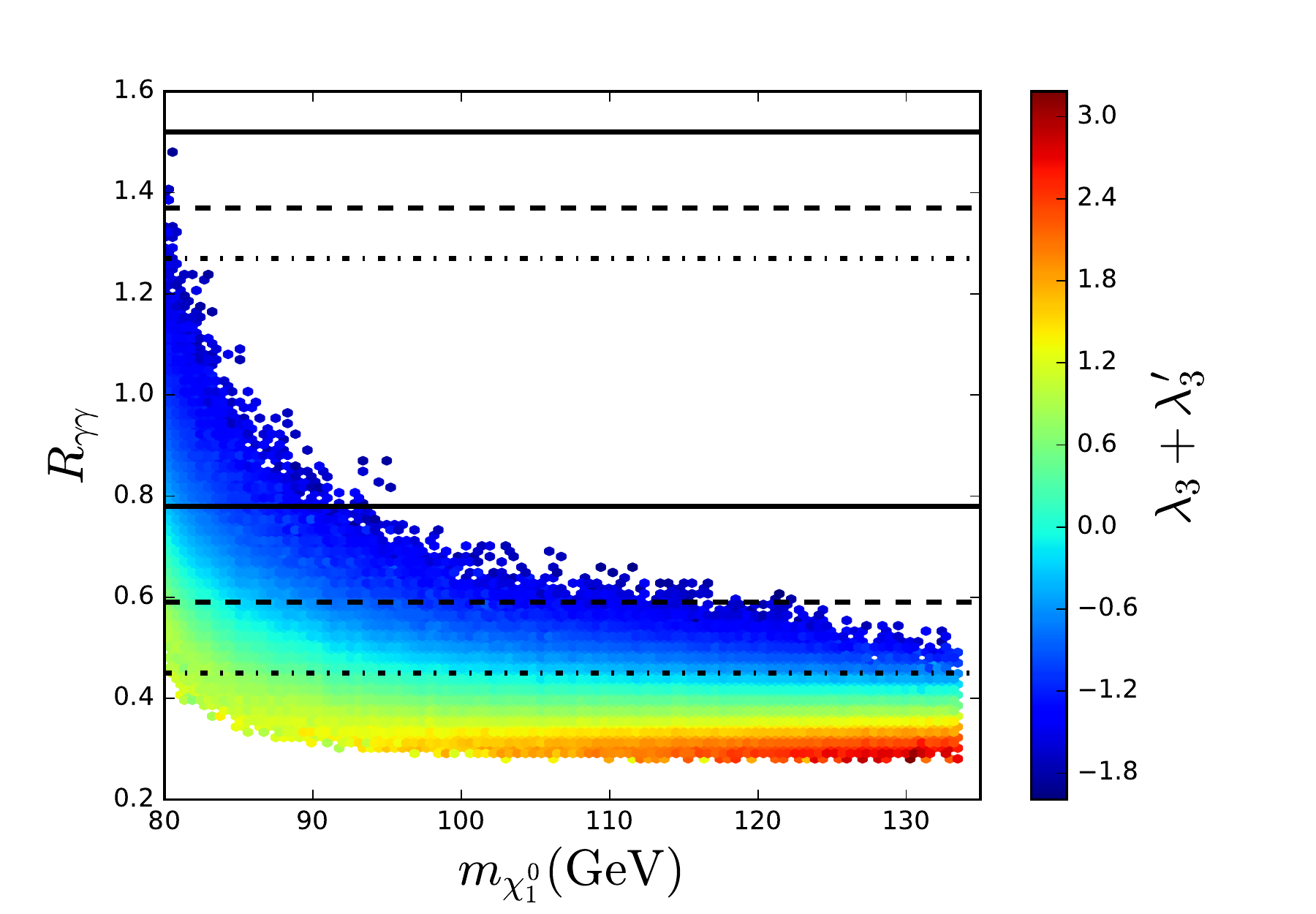}
\includegraphics[scale=0.48]{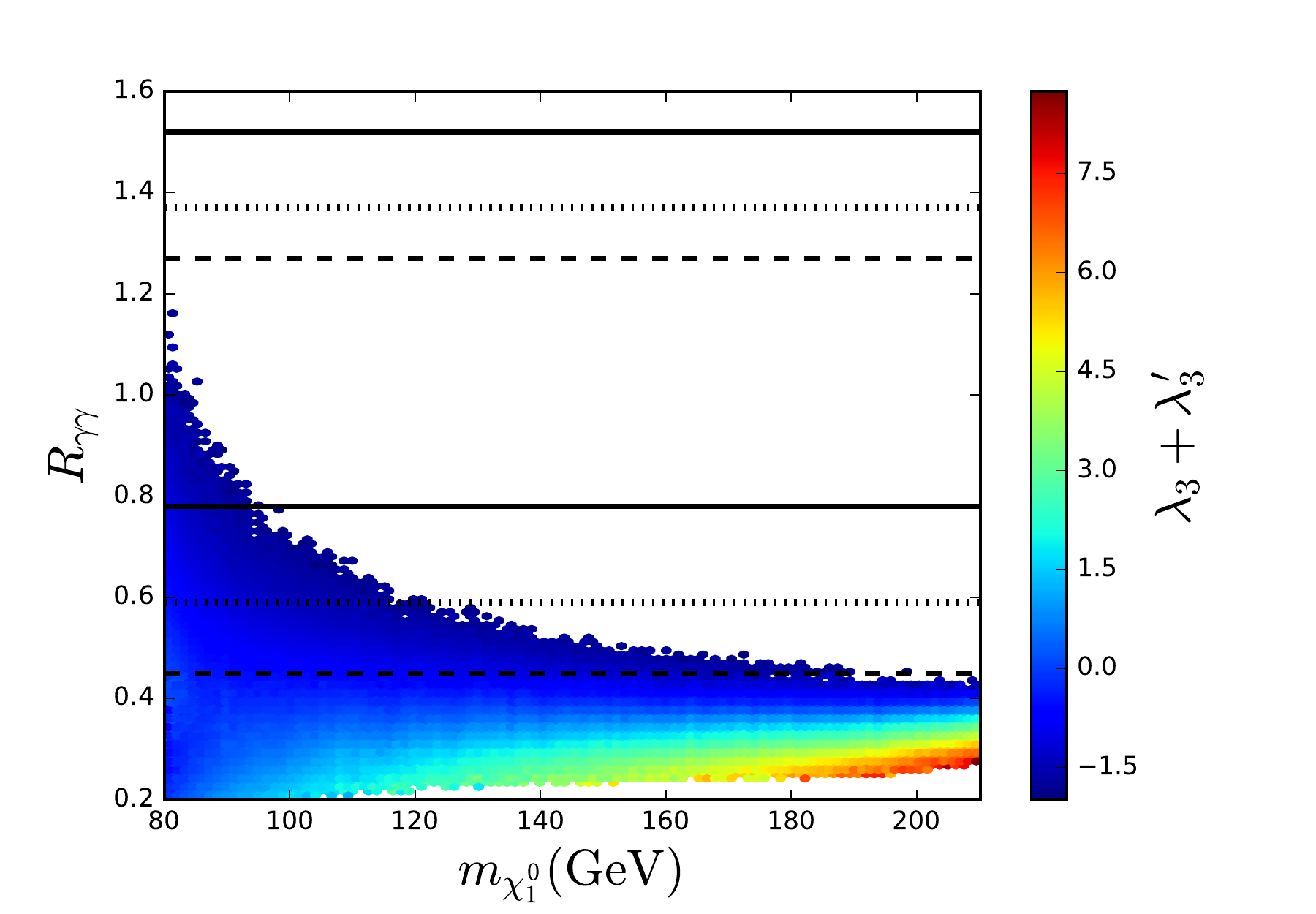}
\caption{The expectation for $R_{\gamma\gamma}$ as a function of the $m_{\chi_1^0}$ in the electroweak DM region, with the color bar displaying the combination of scalar couplings $\lambda_3+\lambda'_3$. The left figure corresponds to the region where $M_{\Sigma} < -m_{\chi_1^0}$ and the right figure to the region where $M_{\Sigma} > -m_{\chi_1^0}$. All points satisfy, additionally, perturbativity, vacuum stability, relic abundance, direct detection, and electroweak precision constraints. The solid, dashed and dotted-dashed lines represent the 2$\sigma$ interval of the Higgs diphoton decay rate from the Run-1 ATLAS+CMS analysis~\cite{Khachatryan:2016vau} and preliminary results from Run-2 reported by CMS~\cite{CMS:2016ixj} and ATLAS~\cite{ATLAS:2016nke}, respectively. }
\label{fig:RggDTsf}
\end{center}
\end{figure}

To explore the impact on $R_{\gamma\gamma}$ we  performed a random scan, taking as input points those that satisfy relic abundance, direct detection constraints, vacuum stability and perturbativity. The relevant parameters are varied as follows:
\begin{align}\label{eq:scan}
& 0<\lambda_2,\lambda_{\Delta}<4\pi/3;\;|\lambda_3|,|\lambda_3'|<4\pi;\; 1.2<m_{{\kappa_1},{\kappa_2}}/m_{\chi_1^0}<3.0;\;\nonumber\\ 
&1.2<m_{\eta_1,\eta_2}/m_{\chi_1^0}<3.0;\; m_{A^0}=m_{\eta_1};\; \mu\ll v.
\end{align}
The last two conditions, together with $m_{\kappa_1} - m_{\eta_1} \leqslant 85$ GeV, which we checked, ensure that the new scalars do not give sizable contributions to the electroweak precision observables, namely the $S$ and $T$ parameters.
The constraint on the mass ratio between the scalar and the DM is imposed to ensure that there are no coannihilation effects; hence, the expected DM phenomenology  is not spoiled.The scan results are shown in Figure~\ref{fig:RggDTsf}, with all the points satisfying the constraints mentioned above. Points between the solid, dashed and dotted-dashed lines are consistent with the Higgs diphoton decay rate from the Run-1 ATLAS+CMS analysis~\cite{Khachatryan:2016vau}, and the preliminary results from the Run-2 ATLAS~\cite{ATLAS:2016nke} and CMS~\cite{CMS:2016ixj} data, respectively. 
It is evident from the plot that $\lambda_3 + \lambda_3'<0$ is required in order to have an $R_{\gamma\gamma}$ within the bounds from the Run-1 ATLAS+CMS analysis~\cite{Khachatryan:2016vau}, with the largest $m_{\chi_1^0}$ at 94 GeV for $M_{\Sigma} < -m_{\chi_1^0}$ and 100 GeV for $M_{\Sigma} > -m_{\chi_1^0}$. Note also that the allowed maximum values of $m_{\chi_1^0}$ tend to be greater if the preliminary results from Run-2~\cite{CMS:2016ixj,ATLAS:2016nke} are taken into account. 

A final discussion is in order on collider bounds which may constrain to some extent the electroweak DM region. The main production processes associated with $Z_2$-odd fermions at the LHC are the same as those of the DTFDM model, $ q\bar{q}'\rightarrow W^{\pm*}\rightarrow \chi_1^{\pm} \chi_2^0$  \cite{Dedes:2014hga,Freitas:2015hsa}. 
Since the $Z_2$-odd scalars are coupled to $Z_2$-odd fermions through the term controlled by $\zeta_i, \rho_i$, and $f_i$, which are assumed small ($\ll1$) in order to be compatible with LFV bounds (see Sec. \ref{sec:neutrinomasses}), it turns out that $\chi_1^{\pm}$ can only decay to $\chi_1^0$ via a virtual $W^\pm$.  Thus, the  characteristic signal to be looked for is events with final states involving three leptons and  missing transverse momentum, again as in the  DTFDM model \cite{Dedes:2014hga,Freitas:2015hsa}.
This signal has been explored by the CMS Collaboration at $\sqrt{s}=8$ TeV \cite{Khachatryan:2014qwa} and $\sqrt{s}=13$ TeV  \cite{CMS:2016gvu}, and by the ATLAS Collaboration \cite{Aad:2014nua}, for the case of supersymmetric charginos and neutralinos, in the limit of $\chi_1^{\pm}, \chi_2^0$ being winolike with decoupled Higgsinos and squarks. When the bounds on the production cross section are translated into this model lesser constraints are obtained \cite{Freitas:2015hsa}, due in part to the nonwinolike character of $\chi_1^{\pm}, \chi_2^0$, which leads to the conclusion that the current LHC results do not constrain the electroweak DM region. 
Nonetheless, the analysis made in Ref.~\cite{Freitas:2015hsa} regarding the future reach of the LHC  shows that this scenario can be completely explored with a luminosity of 300 fb$^{-1}$.  

Regarding collider bounds on the $Z_2$-odd scalars, these must be massive enough ($\gtrsim 45$ GeV) in order to avoid new decay channels of the gauge bosons. 
For the masses considered here ($\gtrsim1.2 m_{\chi_1^0}\approx96$ GeV), such a limit does not apply. 
On the other hand, the main production of the charged scalars at the LHC is via Drell-Yan processes and, for $SU(2)_L$-doublet scalars, their subsequent decays involve gauge interactions and strongly depend on the mass spectrum. The final step is the decay of the lightest $Z_2$-odd particle into the DM particle $\chi_1^0$ plus a lepton through the Yukawa interactions controlled by $\zeta_i$ and $\rho_i$ ($f_i$) for doublet (triplet) scalars. 
Regarding the doublet case and for the mass spectrum  $m_{H^0}, m_{A^0} > m_{H^\pm}$, the decay modes $H^\pm \rightarrow \ell^{\pm} \chi_1^0 $ (with $\ell=e, \mu, \tau$) lead to the collider signature of dileptons plus missing transverse momentum\footnote{The other possible decay modes including virtual gauge bosons increase the number of soft objects or degrade the signal.}. 
This signature is analogous to slepton pair production in the context of simplified supersymmetric scenarios \cite{Aad:2014vma}. 
It follows that assuming a 100\% branching ratio into $\ell=e, \mu$, the analysis done in Ref.~\cite{Hessler:2016kwm} excludes masses up to  $ m_{H^\pm}\sim 160$ GeV. 
However, this limit is loosened either for a degenerate mass spectrum ($m_{H^\pm}- m_{\chi_1^0}\lesssim70$ GeV)~\cite{Aad:2014vma} or when the branching ratio into tau leptons is appreciable or dominant ($\gtrsim$ 0.3)\footnote{Note that a large branching ratio into taus is favoured by the LFV bounds.}. 
For the case of $m_{H^0}< m_{A^0} < m_{H^\pm}$, the decay chain involves the decays $H^\pm\to W^\pm+H^0(A^0)$, $A^0\to Z+H^0$ and $H^0\to \chi_1^0+\nu, \chi_1^\pm+\ell^\mp$, with the resulting signature of multileptons plus missing transverse momentum. 
When $H^0$ directly decays into neutrinos and $\chi_1^0$, the signatures are the same as the ones in the IDM which lead to no applicable bounds on the $Z_2$-odd scalar masses within the electroweak DM region \cite{Belanger:2015kga}. 
Lastly, for the  triplet case we have the single mass spectrum  $m_{\Delta^\pm} > m_{\Delta^0}$, with the decay modes $\Delta^\pm \rightarrow W^{\pm} +\Delta^0 $ and $\Delta^0 \rightarrow \chi_1^0+\nu, \chi_1^\pm+\ell^\mp$. 
As a result, the signatures are multileptons plus missing transverse momentum, for which a similar result with respect to the doublet case is expected. 

\subsection{Scalar DM}

With respect to the DM phenomenology of this model, for the case of a heavy $A^0$ ($m_{A^0}\gtrsim m_{\eta_{1,2}}$), we identify three DM scenarios depending on the size of the entries of the neutral mass matrix: one where the DM is mostly doublet ($(M_{S^0})_{11}\ll(M_{S^0})_{22}$ and $\mu \ll v$), one with triplet DM ($(M_{S^0})_{11}\gg(M_{S^0})_{22}$ and $\mu \ll v$), and a scenario in between ($(M_{S^0})_{11}\sim(M_{S^0})_{22}\sim\mu v$). 
For the first scenario, the  DM particle is $\eta_1\sim H^0$ and the viable DM mass regions are those of the IDM, namely, around the Higgs funnel region and above 500 GeV \cite{Barbieri:2006dq,LopezHonorez:2006gr,Honorez:2010re,LopezHonorez:2010tb,Goudelis:2013uca,Garcia-Cely:2013zga,Arhrib:2013ela,Queiroz:2015utg,Garcia-Cely:2015khw}. 
In the latter region, the so-called high mass regime, the lowest DM mass that reproduces the observed relic DM density is obtained when the interactions of $H^0$ with the Higgs are negligible ($\lambda_{3,4,5}\ll 1$), and so the main channels contributing to DM relic abundance are annihilation of the new charged (neutral) scalars via the $t$- and $u$-channels through the exchange of a neutral (charged) $Z_2$-odd scalar, thus producing gauge bosons. Additionally, the scalar annihilations  mediated by a gauge boson in the $s$-channel are also present. 
As $\lambda_L$ (the parameter that mediates the trilinear interaction of $H^0$ with the Higgs) is increased, new annihilation channels become available, requiring a larger DM mass in order to saturate the relic abundance.  
A similar situation arises for the mostly triplet scenario (where $\eta_2\sim \Delta^0$ is the DM particle): in the pure gauge limit the channels that contribute to the DM relic abundance are similar to those of the mostly doublet case, except for the fact that gauge interactions are now stronger, thus making annihilations more efficient.  As a result, the lowest mass that saturates the relic density is $\sim$ 1.8 TeV \cite{Cirelli:2005uq,FileviezPerez:2008bj,Hambye:2009pw,Araki:2011hm,Lu:2016ucn}.

Regarding the mixed scenario, in the pure gauge limit, the DM particle lies between the above-mentioned cases. If the couplings $\lambda_L$ and $\lambda'_3$ are not zero, the DM mass must be higher to compensate for the larger number of available annihilation channels. Furthermore, in this scenario, if the DM particle is mainly triplet and a degenerate mass spectrum is considered, the constraint on the DM mass would be loosened up to 1.1 TeV due to a net increase in the effective degrees of freedom which lowers the effective cross section. 
On the contrary, when the mass degeneracy within the triplet multiplet is lifted due to the mixing with the doublet (through the $\mu$ term), as well as the doublet components being heavier, it is harder to saturate the relic density constraint due to loss of the effective degrees of freedom entering in the thermally averaged cross section. It is worth noting that this mixed scenario may be significantly constrained from electroweak precision measurements.

For the case of a CP-odd DM particle, $A^0$, we have the same two viable DM mass regions of the IDM, but with the condition of having $\mu\ll v$ in order not to modify the expectations in the high mass regime since it requires the coannihilation of $A^0$ with the other two doublet components.

\section{Neutrino masses}
\label{sec:neutrinomasses}
As it was mentioned above, neither of the doublet-triplet mixed scenarios give an account of neutrino masses. This occurs because the new $Z_2$-odd fields do not couple to the lepton doublet via renormalizable and gauge invariant terms. In other words, in these models the lepton number ($L$) is conserved. Nevertheless, from the combination of these models $L$-violating terms are automatically present, which lead at the end to radiative neutrino masses at the one-loop level. Hereby the interplay of the doublet-triplet scalar and fermion DM models leads automatically to a framework with massive neutrinos. 

\begin{figure}[t!]
\begin{center}
\includegraphics[scale=0.6]{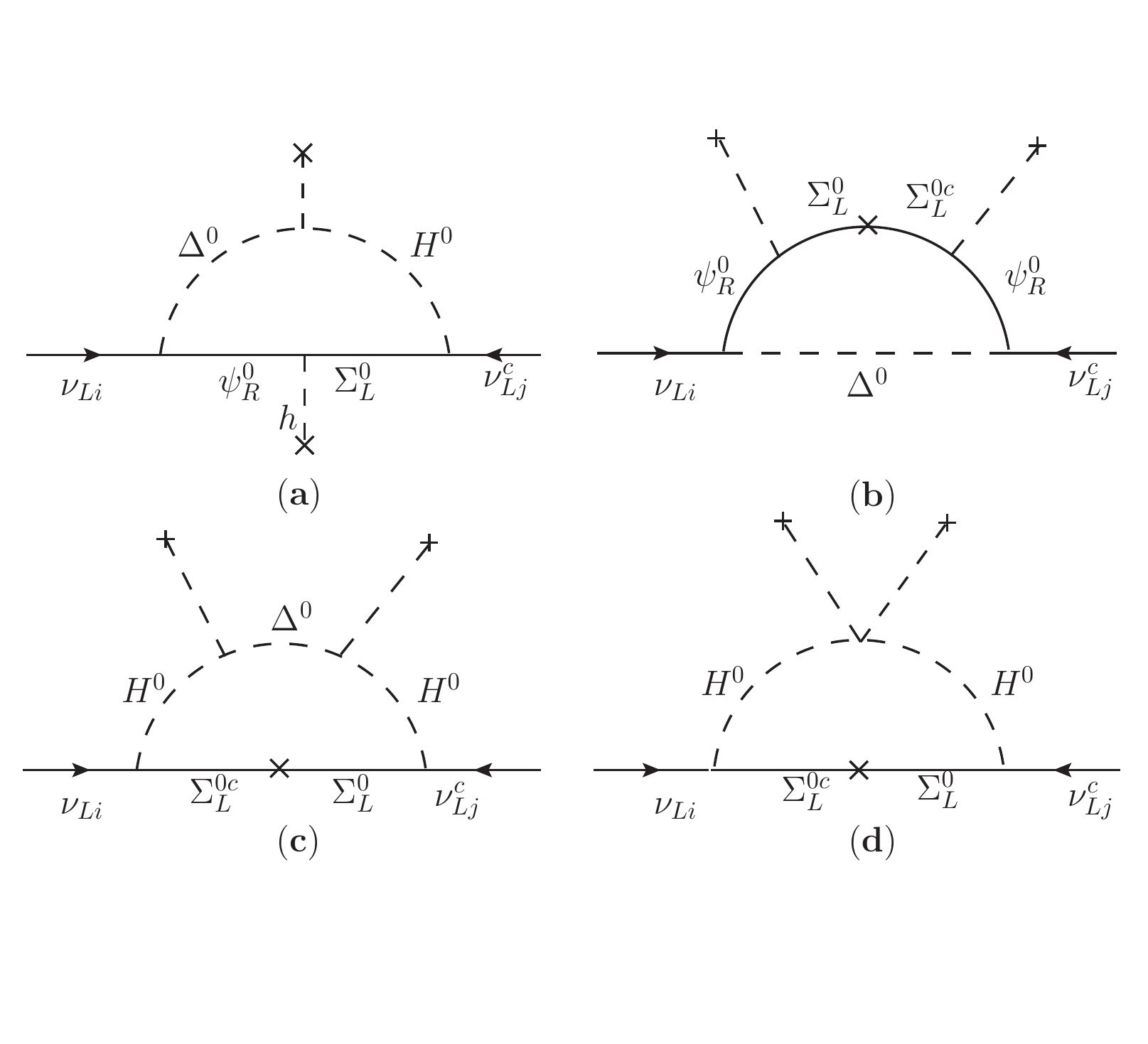}
\vspace{-1.2cm}
\caption{Feynman diagrams leading to one-loop neutrino masses. For the type ({\bf a}) topology there are two Feynman diagrams, one with charged particles running in the loop and one with neutral particles, only the neutral one is shown. }
\label{fig:topologies}
\end{center}
\end{figure}

In the doublet-triplet scalar (fermion) DM model the trivial lepton number assignment $L(H_2)=L(\Delta)=0$ ($L(\Sigma)=L(\psi)=0$) guarantees $L$ conservation. However, when both models are combined several new $L$-violating terms appear. In particular, by keeping the same $L$ assignment the lepton number is violated in one unit by each of the Yukawa terms $\zeta_i$, $f_i$ and $\rho_i$\footnote{In contrast, if we assign lepton numbers for the fermions such that $L(\psi)=1=-L(\Sigma)=1$ and all the new scalars at zero, then lepton number is violated in two units through the $y_1$ term and also in two units by the Majorana-triplet mass term.}.
This in turn means that it is possible to generate neutrino masses at the one-loop level since the seesaw mechanism at tree level is not operative due to the vanishing vev of $H_2$ and $\Delta$. 
Depending on which set of Yukawa couplings are used to build the one-loop neutrino mass diagram we have four different topologies (displayed in Fig. \ref{fig:topologies}) that lead to the three finite realizations of the $d=5$ Weinberg operator \cite{Ma:1998dn, Bonnet:2012kz}. 
Specifically, $\zeta_i$ generate $c)$ and $d)$ diagrams whereas $f_i$ generate a $b)$ diagram, and both set of couplings enter in diagram $a)$. 
In both $a)$ and $b)$ diagrams the mixing fermion term $y_1$ is mandatory and for the $b)$ diagram the mixing scalar term $\mu$ is also required.  
As electroweak eigenstates, the fermion triplet enters in each diagram, the scalar doublet is required in three of them,  
the same occurs for the triplet scalar, and the doublet fermion enters in only two diagrams. 

The general expression for the Majorana mass matrix from all the one-loop contributions displayed in Fig. \ref{fig:topologies} can be written as

\begin{align}
\label{eq:neutrinomass}
M_{\nu}=& \Lambda_{\zeta}\zeta_i \zeta_j+\Lambda_ff_if_j+\Lambda_{f\zeta}(\zeta_if_j+f_i\zeta_j),
\end{align}
where 
\begin{align}
 \label{eq:neutrinomass2}
\Lambda_{\zeta}\;=\;& \frac{1}{32 \pi^2}\frac{1}{2}\sum_{k=1}^3m_{\chi_k^0}\left(U_{1 k}\right)^2\left[ c_{\alpha}^2F_1(m_{\eta_1}^2,m^2_{\chi_{k}^0}) + s_\alpha^2F_1(m_{\eta_2}^2,m^2_{\chi_{k}^0})- F_1(m_{A^0}^2,m^2_{\chi_{k}^0})\right], \nonumber\\
\Lambda_{f}\;=\;&\frac{1}{16 \pi^2}\frac{1}{4}\sum_{k=1}^3 m_{\chi_k^0} \left(U_{3 k}\right)^2 \left[ s_\alpha^2F_2(m_{\eta_1}^2,m^2_{\chi_{k}^0})  + c_\alpha^2F_2(m_{\eta_2}^2,m^2_{\chi_{k}^0})\right], \nonumber\\
\Lambda_{\zeta f}\;=\;& \frac{1}{32 \pi^2}\left[\frac{1}{2}s_\alpha c_\alpha\sum_{k=1}^3m_{\chi_k^0} \ U_{1 k} U_{3 k}\left[ F_1(m_{\eta_2}^2,m^2_{\chi_{k}^0})-F_1(m_{\eta_1}^2,m^2_{\chi_{k}^0}) \right] \right. \nonumber\\
&\left.+s_\theta  c_\theta \displaystyle\sum_{k=1}^2 m_{\chi_k^\pm}\    V^L_{1 k} V^{R*}_{2 k} \left[ F_1(m_{\kappa_1}^2,m^2_{\chi_{k}^\pm})-F_1(m_{\kappa_2}^2,m^2_{\chi_{k}^\pm}) \right]\right],
\end{align}
with $U$, $V^L$ and $V^R$ being the rotation matrices for the neutral, charged-left and charged-right fermions, respectively, 

 and we have used $s_\alpha$ and $c_\phi$ as a short-hand notation for the sine and 
the cosine of a given scalar 
mixing angle $\phi$. The loop functions read
\begin{align}
F_1(m_1^2,m_2^2)=\frac{ m^2_1}{m^2_1-m^2_2} \ln\frac{m^2_1}{m^2_2},\hspace{1cm}
F_2(m_1^2,m_2^2)=\frac{m_1^2\ln m^2_1 - m^2_2 \ln m^2_2}{m^2_1-m^2_2}.
\end{align}
Since $M_\nu$ has a null determinant there is only one Majorana phase, the neutrino spectrum has one massless neutrino, and the two nonzero neutrinos masses are set by the solar and atmospheric mass scales, {\it e.g.}, for normal hierarchy  $m_{\nu_1}=0$, $m_{\nu_2}=\sqrt{\Delta m_{\text{sol}}^2}$ and $m_{\nu_3}=\sqrt{\Delta m_{\text{atm}}^2}$.
Furthermore, it is possible to parametrize five of the six Yukawa couplings in terms of the neutrino observables so that there is only one free parameter. 
Specifically, for the case of a normal hierarchy and by considering without loss of generality $\zeta_1$ as the free parameter, the most general Yuwawa couplings compatible with the neutrino oscillation data are given by
\begin{align}
f_i&=\frac{1}{\Lambda_{f}}\left[\pm\sqrt{\Lambda_{f}\alpha_{ii}-\tilde{\Lambda}\zeta_i^2 } -\Lambda_{\zeta f} \zeta_i\right], \hspace{1cm}i=1,2,3, \\
\zeta_j&= \frac{\pm1}{\Lambda_{f} \tilde{\Lambda}\,\alpha_{11}} \sqrt{ \lambda^2 m_{\nu_2} m_{\nu_3} \Lambda_{f}^2 \tilde{\Lambda}(V_{13}^{*}V_{j2}^{*}-V_{12}^{*}V_{j3}^{*})^2(\Lambda_{f}\alpha_{11}-\tilde{\Lambda}\zeta_1^2 )} + \frac{\alpha_{1j} \zeta_1 }{\alpha_{11}}, \hspace{1cm}j=2,3,
\end{align}
where we have defined
\begin{align}
\tilde{\Lambda}\equiv(\Lambda_{f} \Lambda_{\zeta}-\Lambda_{\zeta f}^2),\,\,\,
\alpha_{ij}\equiv m_{\nu_2} \lambda^2 V_{i2}^* V_{j2}^* + m_{\nu_3} V_{i3}^* V_{j3}^*.
\end{align}

In addition, we have used $\widetilde{M}_\nu=U^{\text{T}}_{\text{PMNS}}\,M_\nu\, U_{\text{PMNS}}$ 

and $U_{\text{PMNS}}=VP$ \cite{Agashe:2014kda}, with $\widetilde{M}_\nu={\rm diag}(0,m_{\nu_2},m_{\nu_3})$, the matrix $V$ containing the Dirac phase and the neutrino mixing angles, and $P=\mbox{diag}(1,\lambda,1)$ giving account of the Majorana neutrino phase.

For the case of an inverted hierarchy the parametrization would yield a similar result which we do not include. In this way, it is always possible to correctly reproduce the neutrino oscillation observables within the DTDM model.

In order to estimate the size of the Yukawa couplings $f_{1,2,3}$ and $\zeta_{2,3}$ for the electroweak DM region, we have repeated the scan over the parameter space (see Eq. (\ref{eq:scan})) with $10^{-3}\leq \zeta_1\leq 1$, and assume CP conservation and a normal hierarchy. As a result we have found that the Yukawa couplings $f_{1,2,3}$ and $\zeta_{2,3}$ can be small as  $\sim 10^{-3}$. Since such couplings also control LFV processes, it follows that the corresponding rates can become rather suppressed because they generically involve the product of two  squared Yukawa couplings. Consequently, in the electroweak DM region it is also possible to be compatible with the LFV constraints \cite{Adam:2013mnn,Aubert:2009ag}.

\section{Conclusions}\label{sec:conclusions}
In this paper we have considered an extension of the SM with an $SU(2)_{L}$ vectorlike doublet fermion, a Majorana triplet, a scalar doublet and a real scalar triplet. Additionally, we imposed a $Z_2$ symmetry which guarantees the DM stability, where all new fields are odd while the SM ones are even. We showed that the model allows for either scalar or fermion DM at the electroweak and TeV scales. 
For fermion DM at the electroweak scale, the DM particle is pure doublet and satisfies the relic density constraint for large Yukawa couplings. We have shown that the most recent limits from direct detection restrict the DM mass to be less than $135$ GeV for $M_{\Sigma}< M_{\psi}$ whereas for $M_{\Sigma}>-M_{\psi}$ the DM mass could go up to $215$ GeV, in both cases with $m_{\chi_1^0}\gtrsim80$ GeV. Additionally we showed that due to the new charged scalar fields, it is possible to satisfy experimental bounds from the LHC Run-1 on the Higgs diphoton decay rate for a DM mass less than $\sim100$ GeV and $\lambda_3 +\lambda_3' \lesssim 0$. The preliminary results from the LHC Run-2 suggest that the allowed DM mass range may be larger. Finally, we found that in this model, Majorana masses for two out of the three active neutrinos are generated at the one-loop level in four different topologies, with all new fields participating in the mass generation mechanism. An expression for the neutrino mass matrix as well 
as a parametrization of the relevant Yukawa couplings in terms of neutrino observables is provided.

\section*{Acknowledgements}
We are thankful to Diego Restrepo, Federico von der Pahlen and Susan Westhoff for enlightening discussions. A. B. has been supported by Colciencias and Universidad EIA Grant No. II12014015, R. L. has been supported by Colciencias, and O. Z. has been partly supported by Sostenibilidad-GFIF and UdeA/CODI Grant No. IN650CE and by COLCIENCIAS through the Grant No. 111-565-84269. A. B. is grateful for the hospitality of the HET group at the University of Florida and acknowledges support from Fulbright.

\bibliographystyle{h-physrev4}
\bibliography{darkmatter}

\end{document}